\documentclass[%
 reprint,
superscriptaddress,
 amsmath,amssymb,
 pre,
]{revtex4-2}

\usepackage{soul}
\usepackage{xcolor}
\usepackage{graphicx}
\usepackage{dcolumn}
\usepackage{amsmath}
\usepackage{mathtools}
\usepackage{bm}

\begin{document}

\title{Symmetric Molecular Dynamics}

\author{Sam Cox}
\affiliation{Department of Chemical Engineering, University of Rochester, Rochester, New York, USA}
\author{Andrew D. White}
\email{andrew.white@rochester.edu, he/him}
\affiliation{Department of Chemical Engineering, University of Rochester, Rochester, New York, USA}

\date{\today}

\begin{abstract}
We derive a formulation of molecular dynamics that generates only symmetric configurations. We implement it for all 2D planar and 3D space groups. An atlas of 2D Lennard-Jones crystals under all planar groups is created with symmetric molecular dynamics. 
\end{abstract}

\maketitle

\section{Introduction}

Molecular dynamics has long been proposed as a method for predicting or understanding crystal structures\cite{parrinello1981polymorphic}. However, any practitioner will confess it is near impossible to observe point group symmetries in molecular dynamics. Here we derive a constraint formulation of molecular dynamics where the symmetry group is an \textit{input}. There is a finite number of symmetry groups. We simply simulate under all symmetry groups to generate symmetric structures.

There are two key ideas to our formulation that correspond to the two components of a space group: the point group symmetry and the Bravais lattice. The point group symmetry is treated as a holonomic constraint. The  constraint equation is a function of positions that is zero when the positions are symmetric. Holonomic constraints are a relatively solved problem and we follow previous approaches\cite{ryckaert1977numerical, andersen1983rattle, tobias1988molecular, ryckaert1985special, edberg1986constrained, ciccotti1986molecular}. The Bravais lattice is a constraint on the simulation lattice vectors that ensures the point group will tile space. Namely, the Bravais lattice specifies the relative lattice vector magnitudes and directions. We ensure our simulations are consistent by working in an unconstrained lattice vector space that is mapped to the correct Bravais lattice via a pre-computed tensor. This frees us to use any NPT method in the unconstrained lattice vector space while still matching the Bravais lattice.

The concept of directly simulating under a symmetry group is unknown to us. The closest examples are methods like symmetry \textit{restraints}\cite{anishkin2010symmetry}. These harmonic restraints generally keep the system close to symmetric, but unlike the method we propose here, no single configuration is actually symmetric. Symmetry has certainly been considered as a \textit{measure} of molecular configurations. For example, \citet{zabrodsky1992continuous} proposed a continuous symmetry measure, which is used to quantify the symmetry of atoms. This has been used to directly optimize Lennard-Jones clusters with symmetry\cite{oakley2013symmetrisation}. Of course, the direct use of symmetry for crystal structure prediction with Monte Carlo is common\cite{van1999upack, fredericks2021pyxtal}, and generative models with explicitly included symmetry are common\cite{xie2021crystal}. There are no molecular dynamics methods though that can directly sample space groups, which would be useful for crystal structure prediction and modeling biological assemblies\cite{cannon2019high}. Symmetric molecular dynamics may also be viewed as a special case of objective molecular dynamics, which is a general method that encompasses any infinite or finite periodic tiling of a simulation\cite{dayal2010nonequilibrium, xu2019collapsed}. Similarily, others have explored generalizing periodic boundary conditions to other tilings\cite{hansson2002molecular,denton2009symmetry, wagner2004molecular,roy2011microscopic}. 

Below we derive our equations of motion and discuss implementation details. To assess the method, we show that it conserves energy and is capable of working in arbitrary space groups. Then we demonstrate its use to enumerate crystal structures of the Lennard-Jones potential under all planar groups with NPT simulations. 

\section{Theory}

\subsection{Equation of Motion}

Consider the dynamics of $N$ indistinguishable particles in $D$ dimensions under a Hamiltonian $H(\mathbf{p}(t), \mathbf{q}(t))$. We wish to constrain $H$ so that $\mathbf{q}(t)$ is symmetric at all times. Symmetry is a property of $\mathbf{q}(t)$ and a specific symmetry group of position transformations $G$, like mirror along the $x$ axis. $\mathbf{q}(t)$ is point group symmetric if applying any element of the group results in no change to the positions (ignoring ordering of particles):

\begin{equation}
\label{eq:symm}
    g\cdot\mathbf{q}(t) \sim \mathbf{q}(t),  \forall g \in G
\end{equation}

where $g\,\cdot$ means applying the group element to each particle individually, $\sim$ means row equivalence, and $G$ is a finite group. Group elements are represented as affine matrices in space and planar groups.

Equation~\ref{eq:symm} may hold trivially. For example, all particles are at the origin. Such special positions that are invariant to group elements are known as special Wyckoff positions\cite{wyckoff1922analytical}. We remove this assumption in Section~\ref{sec:wyckoff}, but for now additionally assume

\begin{equation}
    \label{eq:gencond}
    g\cdot\mathbf{q}_i(t) = \mathbf{q}_i(t)\quad \textrm{iff} \, g = I
\end{equation}
where $I$ is the identity transformation.

Assuming Equations~\ref{eq:symm} and \ref{eq:gencond} hold at $t = 0$, the particles can be partitioned into $N / |G| = n$ group orbits. A group orbit is the set generated by applying all elements of group $G$ to positions $\mathbf{q}_i(t)$

\begin{equation}
    \label{eq:orbit}
    G\left[\mathbf{q}_i(t)\right]= \,\{g_j\cdot \mathbf{q}_i(t), g_j \in G\}
\end{equation} 

One member of all orbits will be $\mathbf{q}_i(t)$ itself, because $G$ contains the identity element. We can label the particles as $q_{ij}(t)$ where $i$ indicates the orbit and $j$ indicates the group element. In crystallography the $\mathbf{q}_{i0}$ particles are called the asymmetric unit.  We can satisfy Equation~\ref{eq:symm} at all $t$ by specifying the following holonomic constraint:

\begin{equation}
    \label{eq:constraint}
    \sigma(\mathbf{q}_{ij}) = g_{j}\cdot \mathbf{q}_{i0} - \mathbf{q}_{ij} = \mathbf{0}
\end{equation}

There are $|G| - 1$ of these constraints per group orbit and each removes $D$ degrees of freedom. This means the degrees of freedom of the dynamics is $D\times (n - 1)$.  We can simulate dynamics under the holonomic constraints by simply only modeling the asymmetric unit --- they are the generalized coordinates.\footnote{You can also arrive at this solution by constraining the Cartesian coordinates and finding Lagrange multipliers.}

Thus, our algorithm is to only integrate the asymmetric unit and explicitly consider the remaining $(N - n)$ particles only when computing forces. This is similar to Dayal et al. \cite{dayal2010nonequilibrium}. Practically this is done by setting these constrained particles positions just before computing forces. Similar to work on periodic boundary conditions, these equations of motion may lead to linear momentum conservation problems\cite{shirts2006periodic, kuzkin2015angular}.

One feature of nearly all potentials used in molecular dynamics is that they are $G$-invariant, where $G$ is any planar, space, or permutation group: $U(g\cdot \mathbf{q}) = U(\mathbf{q})$. That makes the forces, $F(\mathbf{q})$, $G$-equivariant:

\begin{equation}
    \label{eq:equiv}
    F(g\cdot \mathbf{q}) = -g\cdot\nabla U(\mathbf{q})
\end{equation}

because the potentials are composed of angles and distances, which are invariant to rotations, mirrors, and permutations\cite{musil2021physics,whiteDeep2021}. For a pairwise potential, we can use Equations~\ref{eq:orbit} and \ref{eq:equiv} to rewrite the potential as
\[
     U(\mathbf{q}) = \sum_{i<j} u\left(\mathbf{q}_i - \mathbf{q}_j\right)
     \]
     \begin{equation}
     = \frac{|G|}{2}\sum_{i=0}^n\sum_{j=1}^{|G|}  u\left(\mathbf{q}_{i0} - \mathbf{q}_{ij}\right)+ |G|\sum_{i=0}^n\sum_{k=i+1}^n\sum_{j=0}^{|G|} u\left(\mathbf{q}_{i0} - \mathbf{q}_{kj}\right)
\end{equation}

where the $|G|$ factor accounts for intra-group orbit interactions that are not explicitly computed. This translates an algorithm of an outer loop over the asymmetric unit and an inner loop over all particles. 

\subsection{Bravais Lattice}

A space group consists of both a point group and a Bravais lattice. The Bravais lattice is specified with $D$ $D$-dimensional unit cell vectors. Particles always remain in one cell among the lattice cells, which are called images. For example, we could simulate the ``root'' cell and its 26 neighboring cells in 3 dimensions. We follow the approach above and treat each image of the system with virtual particles while only integrating the root cell. This means all images of the system are explicit and we can violate the minimum image convention. We were not signatories of the minimum image convention anyway. This approach allows the cell vectors to shrink well below the distance cut-off of the potential, provided we have enough virtual particles to populate past the cut-off of the asymmetric unit of the origin cell. You can simulate $3^{aD}$ images to allow the cells to shrink to at least $1/a$ the cut-off distance.

We need to convert between the fractional coordinates, which are used to tile the particles and apply the point group symmetry, to the Cartesian coordinates, which are used for integration and computed potentials. Given the box vectors in row-form $B$, we can transform between the representations via

\begin{equation}
    \label{eq:scale}
    \mathbf{s}(t) = B^{-1}\mathbf{q}(t),\quad
    \mathbf{q}(t) = B\mathbf{s}(t)
\end{equation}
where $s(t)$ is the fraction of each lattice vector (i.e., fractional coordinates). Wrapping is trivial with fractional coordinates: $s(t) \,\textrm{fmod}\, 1.0$ will wrap the coordinates. All point group transformations are applied in $\mathbf{s}(t)$, however a $B^{-1}$ term should be added to Equation~\ref{eq:orbit} so that it operates on fractional coordinates.

Bravais lattices include more than just the usual cubic and triclinic lattices commonly seen in molecular dynamics barostats. To ensure the cell vectors are consistent with the Bravais lattice while changing box size, we define a tensor $\mathbf{L}$ of shape $D \times D \times D \times D$ that maps from a triclinic box vectors to the proper Bravais Lattice box vectors of the space group. For example, $L_{2011}$ is the contribution to Bravais lattice vector $2$'s $x$ component from triclinic box vector $1$'s $y$ component.  There are many choices that could be made for $\mathbf{L}$. For example, to make a cubic Bravais lattice from a triclinic box vector we require a single parameter $a$ to define the three lattice vectors $(a,0,0), (0,a,0), (0,0,a)$. We could set $a$ by averaging all the vector lengths, averaging all vector components, or select $a$ to be the first element of the first vector. Each of these choices gives a different $\mathbf{L}$ and some have large null spaces. NPT is then accomplished via scaling Monte Carlo moves in the triclinic box vectors ($B'$) following \citet{frenkel1982monte}, and the proper Bravais Lattice is computed via $B = \mathbf{L}B'$.

\subsection{Wyckoff Positions}
\label{sec:wyckoff}

It is possible to have particles that violate Equation~$\ref{eq:gencond}$ while still satisfying Equation~\ref{eq:symm} if $q_{i0}$ is in a special position called a Wyckoff position --- like the origin\cite{wyckoff1922analytical}. To perform constrained molecular dynamics of particle $q_{i0}(t)$ occupying a Wyckoff position, we define a subgroup $G'$ that contains the elements of $G$ which do not leave $q_{i0}(t)$ invariant plus an identity group element. The identity of this subgroup is not the identity transform, but instead a transform that projects from a general position into the Wyckoff position. For example, the Wyckoff position may be the vertical line $x = 0$ and the identity group element would be the transform $x' = 0, y' = y$. We will denote this group element as $P$ to hint it is a projection. 

The group orbit is similarly defined on the subgroup and the other procedures above apply. However, $q_{i0}(t)$ must stay in a Wyckoff position at all time to satisfy Equation~\ref{eq:symm}. This can be accomplished via traditional constrained molecular dynamics of Lagrange multipliers \cite{miyamoto1992settle}. Omitting the indices on $\mathbf{q}_{i0}(t)$, our holonomic constraint is:

\begin{equation}
    \label{eq:wconstraint}
    \sigma\left[\mathbf{q}(t)\right] = P\mathbf{q}(t) - \mathbf{q}(t) = \mathbf{0}
\end{equation}
and the force from the constraint will be
\begin{equation}
    \mathbf{F_c} = \mathbf{\lambda} J[\sigma] = \mathbf{\lambda}\left(P - I\right)
\end{equation}
where $J[\sigma]$ is the Jacobian of $\sigma$ with respect to constraint dimension and element of $\mathbf{q}(t)$. We can solve for $\mathbf{\lambda}$ by knowing that $\sigma\left[\mathbf{q}(t + \Delta t)\right] = \mathbf{0}$:

\begin{equation}
    \label{eq:final_constraint}
    \mathbf{\lambda} = \frac{\Delta t^2}{m}\left[(P - I)^2\right]^{-1}    \sigma\left[\mathbf{q}\,'(t + \Delta t)\right] 
\end{equation}
where $\Delta t$ is the timestep, $m$ is the mass of the particle, and $\mathbf{q}\,'(t + \Delta t)$ is $\mathbf{q}(t)$ integrated without the constraint force by $\Delta t$. All terms are constant except $\sigma\left[\mathbf{q}\,'(t + \Delta t)\right]$, which simplifies computation. 

\begin{figure}
    \centering
    \includegraphics[height=\columnwidth]{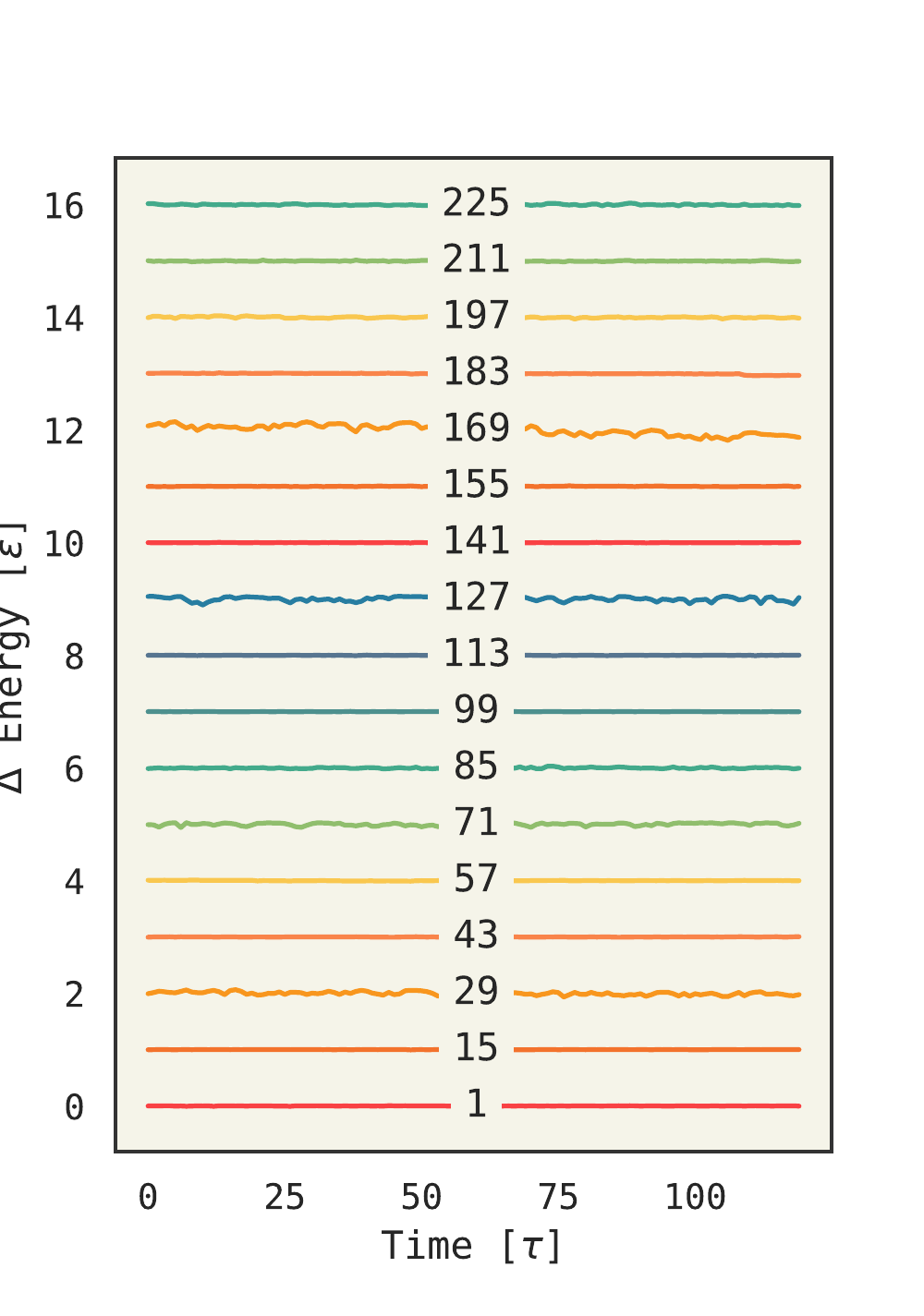}
    \caption{Total energy from NVE simulations under different symmetry groups in 3D. Groups are indicated with Hall numbers. 4 particles are in the asymmetric unit, and the simulations are at a number density of 0.2 and starting temperature of 0.5. The increase of fluctuations is because the unit cell (total particles) increases with size of symmetry group.}
    \label{fig:energy3d}
\end{figure}

\begin{figure*}
    \centering
    \includegraphics[width=\textwidth]{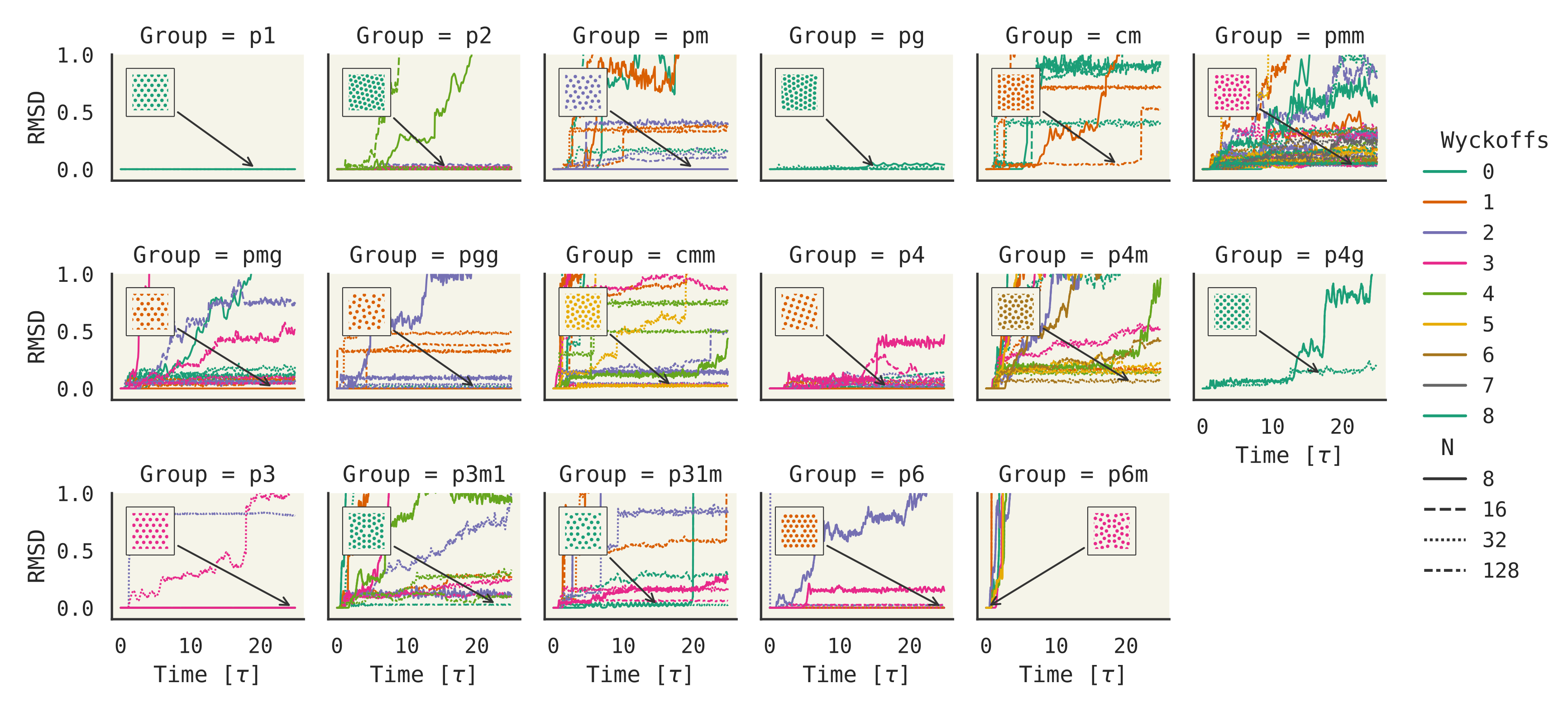}
    \caption{An atlas of 258 Lennard-Jones crystal structures. Subplots are broken down by planar group and each line color/style indicates Wyckoff occupancy and unit cell particle count. Individual plots show RMSD from starting configuration, which was constrained to match the planar group in subplot titles. RMSD was calculated during NVE simulation of 5k steps with no symmetry constraints. A rise indicates change of configuration -- meaning the starting configuration was not stable. Missing traces indicate either simulation diverged or number density did not exceed 0.5.}
    \label{fig:atlas2d}
\end{figure*}

\section{Methods}

We use the BAOAB Langevin dynamics integrator described in \cite{leimkuhler2013robust, leimkuhler2016efficient}. Equation~\ref{eq:constraint} is applied during position updates and Equation~\ref{eq:orbit} is applied before velocity updates (force computation). Degrees of freedom is computed from number of asymmetric unit particles and deducted degrees of freedom from Wyckoff position restraints. All simulations are Lennard-Jones potentials with cutoff 3.5 and in reduced units. NVE simulations are conducted with the velocity-verlet integrator. A timestep of 0.005 and a Langvin $\gamma$ of 0.1 were used for all simulations. Since images are explicit in our implementation, it is necessary to specify the number. We use an image radius of 2 -- meaning $3^{2D}$ images are simulated where $D$ is the dimension. To generate starting configurations, points were randomly generated and filtered to fit into the space group asymmetric unit as specified by \citet{aroyo2013international}. Point group generators and Wyckoff sites were taken from Bilbao crystallography server\cite{aroyo2011crystallography, aroyo2006bilbao, aroyo2006bilbao2}.

We define our results in reduced units, as defined in reference \cite{frenkel1996understanding}. Specifically, energy ($\epsilon$) is fundamental and $\tau$ is a derived unit of the form:
\begin{equation}
\label{eq:red_un}
    \tau = L\sqrt{\frac{m}{\epsilon}}
\end{equation}
where L, m, $\epsilon$ are the fundamental units of length, mass, and energy, respectively.

\section{Results}

We first consider if our implementation conserves energy. Figure~\ref{fig:energy3d} shows the total energy of NVE simulations under a subset of space groups with 5 particles in the asymmetric unit. These were done at number densities of 0.2, with starting temperature 0.5, and for 30k timesteps. The bottom trace (P1) has no symmetry constraints and shows good conservation. There are more fluctuations at other symmetry groups because there are more particles in their unit cells and thus higher energy fluctuations. For example, space group 127 (P4/mbm) has 80 particles in a unit cell when there are 5 in the asymmetric unit, meaning the interaction potential felt has more particles contributing to it. 

Figure~\ref{fig:atlas2d} shows an enumeration of crystal structures under different symmetry groups for a 2D Lennard-Jones fluid. The structures are generated in 2 steps. First, we simulate under a symmetry group constraint in NPT ($P=0.25,\, T=0.1$) for 1M steps. Next, we do a constrained equilibration under NVT for 100k steps at $T = 0.05$. This structure is then the proposed crystal structure for the given symmetry group. Figure~\ref{fig:atlas2d} shows the root mean squared deviation (RMSD) if the resultant structure is simulated under no symmetry constraint in NVE for 5k steps. The assumption is that if the structure is does not collapse (RMSD rise), it is meta-stable. We indeed find that this protocol under no symmetry constraints (p1) gives the correct hexagonal packing. 

To enumerate all planar groups in 2D, we simulate under each group, with 4 choices of particle number (in unit cell), and varying occupancy of Wyckoff sites. As expected, the planar groups with hexagonal Bravais lattices (or permit them) have stable structures: p1, p2, pg, p3, p6, and cmm\footnote{This depends on Wyckoff site occupancy too}. Some unusual stable structures are seen without hexagonal close-packing like p4m and p3m1. Meta-stable structures like these would be nearly impossible to generate without symmetry constraints in molecular dynamics. We find some symmetries have no meta-stable structures: p6m and most square close-packing (cm, cmm, p4g). Interestingly, voids seem to be the way to stabilize these close-packing structures like in pmg. 

\section{Discussion}
\label{sec:discussion}

Here is our advice on implementing symmetric molecular dynamics in a modern molecular dynamics engine. The asymmetric unit should be integrated as usual. Make the non-asymmetric unit particles (images) be ghost particles;  ghost particles are non-integrated particles used in force-field calculations. The ghost particles' positions should be set using affine matrices defining the group transformations in fractional coordinates. These matrices can be obtained from our library or crystallography tables.

Pressure computed for the asymmetric unit is not meaningful and NPT should be done using the algorithm described above that does not require computing pressure from a virial. The lattice vectors may be stored separately than the usual lattice vectors and are only used to set the ghost particle positions.  The tensor transforms can be loaded from our library. Periodic boundary conditions should be disabled entirely if doing NPT.

The constraints for Wyckoff sites are implemented as Lagrange multiplier constraints. The terms can be computed analytically at each step using Equation~\ref{eq:final_constraint}.

\section{Conclusions}

We have formulated a symmetric molecular dynamics algorithm and implemented it. Results show that it can do NPT to enumerate meta-stable crystal structures. A reference implementation is available at \texttt{https://github.com/whitead/symd} and implementation notes for modern simulation is given in the discussion section.

\section{Acknowledgements}

We thank Prof. Glen Hocky and Dr. Charles Matthews for valuable discussion and feedback. This work was supported by the NSF under grant 1751471.

\bibliography{bibliography}

\end{document}